\documentclass[Conference]{IEEEtran}

\usepackage{url}
\usepackage{graphicx}
\usepackage{subcaption}
\usepackage{amsfonts}
\usepackage{multicol}
\usepackage{booktabs}
\usepackage{textcomp}
\usepackage{comment}
\usepackage{mathtools}
\usepackage{stfloats}
\usepackage[below]{placeins}
\usepackage{cite}
\usepackage{pifont}
\usepackage{amssymb}
\usepackage{multirow}
\usepackage{amsmath}
\usepackage{csquotes}
\usepackage{color,soul}
\usepackage[bookmarks=false]{hyperref}
\usepackage{algorithm}
\usepackage{algorithmic}
\usepackage[flushleft]{threeparttable} 

\newcommand{\floor}[1]{\left\lfloor #1 \right\rfloor}

\ifCLASSINFOpdf
\else
\fi

\begin{document}

\title{FPGA Implementation of Stair Matrix based Massive MIMO Detection}

\author{\IEEEauthorblockN{Shahriar Shahabuddin\IEEEauthorrefmark{1},
Mahmoud~A.~Albreem\IEEEauthorrefmark{1}\IEEEauthorrefmark{3},
Mohammad Shahanewaz Shahabuddin\IEEEauthorrefmark{2},\\
Zaheer Khan\IEEEauthorrefmark{1},
Markku Juntti\IEEEauthorrefmark{1}
\\}

\IEEEauthorblockA{\IEEEauthorrefmark{1}Centre for Wireless Communications, University of Oulu, Finland\\
\IEEEauthorrefmark{3}A'Sharqiyah University, Oman\\
\IEEEauthorrefmark{2}Vaasa University of Applied Science, Finland\\
}

\IEEEauthorblockA{ Email: \IEEEauthorrefmark{1}[firstname.lastname]@oulu.fi,   
\IEEEauthorrefmark{2}e1900302@vamk.fi,
\IEEEauthorrefmark{3}[firstname.lastname]@asu.edu.om}  
}

\maketitle

\begin{abstract}

Approximate matrix inversion based methods is widely used for linear massive multiple-input multiple-output (MIMO) received symbol vector detection. Such detectors typically utilize the diagonally dominant channel matrix of a massive MIMO system. Instead of diagonal matrix, a stair matrix can be utilized to improve the error-rate performance of a massive MIMO detector. In this paper, we present very large-scale integration (VLSI) architecture and field programmable gate array (FPGA) implementation of a stair matrix based iterative detection algorithm. The architecture supports a base station with 128 antennas, 8 users with single antenna, and 256 quadrature amplitude modulation (QAM). The stair matrix based detector can deliver a 142.34 Mbps data rate and reach a clock frequency of 258 MHz in a Xilinx Virtex-7 FPGA. The detector provides superior error-rate performance and higher scaled throughput than most contemporary massive MIMO detectors.
\end{abstract}


\section{Introduction}\label{intro}

Massive multiple-input multiple-output (MIMO) is a key technology for fifth generation (5G) systems to enhance spectral and energy efficiency, coverage and mobility within the available radio spectrum. Massive MIMO is the successor of conventional small-scale MIMO, where the number of antennas at the base station (BS) and the number of users is relatively high~\cite{MIMO_survey}. Despite all the benefits of massive MIMO, the technology also suffers from higher computational complexity. MIMO detection in the uplink is one of the most complex part of a massive MIMO BS due to the increasing number of users and BS antennas. The complexity of massive MIMO detectors increase so rapidly that conventional exact inversion-based linear massive MIMO detection might be too complex for certain BS products. In consequence, a new class of detectors based on approximate inversion based methods have become a popular choice among the large-scale integration (VLSI) community over the past decade. Approximate inversion based detectors are based on the principle that for certain antenna configurations, the equalization matrix of linear detection is diagonally dominant. Several approximate inversion based detectors, which utilize a diagonal matrix can be found in the literature, such as 
Neumann series approximation (NSA)~\cite{NSA_Chris_Rice}, Gauss Seidel (GS)~\cite{chuan_zhang_GS}, conjugate gradient (CG)~\cite{MatDecomposition_mMIMO}, Richardson method~\cite{Richardson} etc. 

Instead of diagonal matrix, a stair matrix can be utilized for massive MIMO detection~\cite{stair_matrix1}. With a small additional complexity, the stair matrix based detectors can accelerate convergence rate of an approximate inversion based detectors, such as NSA. In addition, a stair matrix can also be used to develop novel low-complexity massive MIMO detection methods. 
In this paper, we propose FPGA implementation for such a stair matrix based detection method. We present simulation results to demonstrate superior error-rate performance of stair matrix based detection and also determine necessary word length for the circuit. We develop an iterative and time-shared architecture for the detector using VHSIC Hardware Description Language (VHDL) and mapped on a Xilinx Virtex-7 FPGA. The architecture supports a base station with 128 antennas, 8 users and 256 quadrature amplitude modulation (QAM). The stair matrix based detector provides 142.34 Mbps detection rate at a 258 MHz clock frequency. The rest of the paper is organized in the following way: In Section~\ref{system}, a massive MIMO system and detection methods are discussed. In Section~\ref{stair}, the stair matrix based detection algorithm and its complexity is presented. Fixed and floating-point performance of the detector are presented in Section~\ref{simulation}. Proposed VLSI architecture and FPGA implementation are presented in Section~\ref{vlsi} and~\ref{fpga}, respectively. The conclusion is drawn in Section~\ref{Conclusion}. 

\section{System Model and Detection Methods}\label{system}

\begin{figure}[h] 
\centering
\includegraphics[keepaspectratio,width=.8\columnwidth]{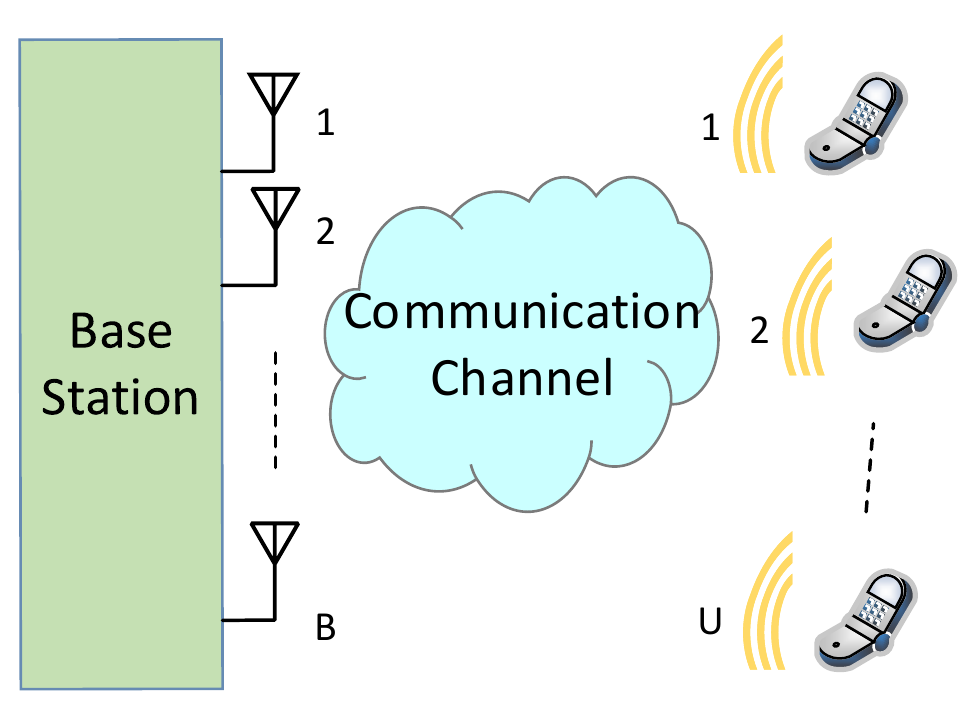}
\caption{Massive MIMO system model.}
\label{fig:smodel}
\end{figure}

We assume, a total of $U$ single antenna users are transmitting towards a massive MIMO BS with $N$ antennas, where $U\leq N$. Assuming a frequency flat channel between the users and BS, the transmit and receive vector relationship can be characterized as
\begin{equation}\label{gf}
    \mathbf{y}= \mathbf{H} \mathbf{x}+\mathbf{n},
\end{equation}
where $\mathbf{y} \in \mathbb{C}^{B}$ is a received signal vector, $\mathbf{x} \in \mathbb{C}^{U}$ is a transmit symbol vector, $\mathbf{H} \in\mathbb{C}^{B \times U}$ is a channel matrix, and $\mathbf{n} \in \mathbb{C}^{B}$ is a circularly symmetric complex white Gaussian noise vector with zero mean and $\sigma^2$ noise variance. In Fig.~\ref{fig:smodel}, a massive MIMO system model is presented.

The task of a MIMO symbol detector is to determine the transmitted symbol vector $\mathbf{x}$ from the received signal vector $\mathbf{y}$. Two recurrent linear detection methods are based on zero-forcing (ZF) and linear minimum mean-square error (MMSE) equalization. ZF inverts the channel matrix $\mathbf{H}$ to determine the transmitted vector and does not consider the effect of noise vector $\mathbf{n}$, which can be expressed as
\begin{equation}
\mathbf{\tilde{x}}_{\text{ZF}}= \mathbf{H}^{\dagger}\mathbf{y}=(\mathbf{H}^\mathbf{H}\mathbf{H})^{-1}\mathbf{H}^\mathbf{H}\mathbf{y},
\end{equation}
where $\mathbf{H}^{\dagger}$ is a pseudo-inverse of $\mathbf{H}$. The ZF detector requires an inversion of the \textit{Gramian} matrix, $\mathbf{G}$, where $\mathbf{G}_{\text{ZF}}=\mathbf{H}^\mathbf{H}\mathbf{H}$. 
The MMSE detector takes noise into account and provides better performance than ZF. MMSE detection can be expressed as
\begin{equation}
\mathbf{\tilde{x}}_{\text{MMSE}}=(\mathbf{H}^\mathbf{H}\mathbf{H}+\sigma^2\mathbf{I}_U)^{-1}\mathbf{H}^\mathbf{H}\mathbf{y},
\end{equation}
where $\mathbf{I}_U$ is the $U \times U$ identity matrix. The \textit{Gramian} matrix is modified with a regularization by noise variance for MMSE, i.e., $\mathbf{G}_{\text{MMSE}}=\mathbf{H}^\mathbf{H}\mathbf{H}+\sigma^2\mathbf{I}_U$.

For small-scale MIMO detection, the inversion of a \textit{Gramian} was based on exact matrix inversion methods. However, the exact inversion can be complex as the number of users increases. For example, \textit{Gramian} of a 16 users system ($U=16$) will be a $16\times16$ matrix. Therefore, the approximate inversion based detection methods are usually used by the VLSI community. These detection mechanisms are iterative and they heavily utilize the diagonal matrix. For example, in GS method, $\mathbf{G}$ is decomposed as 
\begin{equation}
\mathbf{G} = \mathbf{D} + \mathbf{L} + \mathbf{R},
\end{equation}
where $\mathbf{D}$, $\mathbf{L}$ and $\mathbf{R}$ are the diagonal component, the strictly lower triangular component, and strictly upper triangular component, respectively. The GS can be used to estimate the transmitted signal vector $\hat{\mathbf{x}}$ as
\begin{equation}
\hat{\mathbf{x}}_{t} = \left ( \mathbf{D} + \mathbf{L} \right )^{-1}\left ( \hat{\mathbf{x}}_{\text{MF}} - \mathbf{R}\hat{\mathbf{x}}_{t-1} \right ),
\label{Eq. 22}
\end{equation}
where $\hat{\mathbf{x}}_{\text{MF}}=\mathbf{H}^H\mathbf{y}$ is a matched filter~\cite{chuan_zhang_GS}.

\section{Stair Matrix based Gauss-Seidel Method} \label{stair}

A stair matrix is a special tri-diagonal matrix where the off-diagonal elements on either the even or the odd row are zeros \cite{stair_matrix1}. Matrix $\mathbf{S}$ is considered as a stair matrix if one of the following conditions is satisfied:

\begin{itemize}
\item[-] $\mathbf{S}_{\left ( i,i-1 \right )} = 0, \mathbf{S}_{\left ( i,i+1 \right )} = 0$ where $i = 2,4,..., 2\left \lfloor \frac{K}{2} \right \rfloor$
\item[-] $\mathbf{S}_{\left ( i,i-1 \right )} = 0, \mathbf{S}_{\left ( i,i+1 \right )} = 0$ where $i = 1,3,..., 2\left \lfloor \frac{K-1}{2} \right \rfloor+1$
\end{itemize}

The stair matrix $\mathbf{S}$ is essentially a diagonal matrix $\mathbf{D}$ with some additional off-diagonal elements. A $6 \times 6$ stair matrix can be expressed with either of the following forms:

\scalebox{0.83}{%
$\mathbf{S} = \begin{bmatrix}
\times & \times &  &  &  & \\ 
 &  \times&  &  &  & \\ 
 &  \times& \times & \times &  & \\ 
 &  &  & \times &  & \\ 
 &  &  & \times & \times &\times \\ 
 &  &  &  &  & \times
\end{bmatrix} 
\quad \text{or} \quad \mathbf{S} = \begin{bmatrix}
 \times &  &  &  &  & \\ 
 \times & \times & \times &  &  & \\ 
 &  & \times & &  & \\ 
 &  & \times & \times & \times & \\ 
 &  &  &  & \times & \\ 
 &  &  &  & \times & \times
\end{bmatrix}$
}

The stair matrix $\mathbf{S}$ can be extracted from the \textit{Gramian} by only extracting the cross shaped values shown here. The inversion of a stair matrix can be calculated in a straightforward manner, which is presented in Algorithm \ref{algo1}.
The diagonal elements are computed with reciprocals and the off-diagonal elements require a couple of multiplications.
\begin{algorithm}[t]
	\caption{Inversion of a Stair Matrix}
	\label{algo1}
	\begin{algorithmic}
		\STATE{\textbf{\textit{input}}: $\mathbf{S}$}
	    \STATE{\textbf{\textit{outputs}}: $\mathbf{S}^{-1}$ }
		\STATE{1: $\textbf{for}$ $i=1:1:U$}
		\STATE{2: \hspace{4 mm} $\mathbf{S}^{-1}_{i,i} = 1/\mathbf{S}_{i,i} $}
		\STATE{3: $\textbf{end}$}
		\STATE{4: $\textbf{for}$ $i=2:2:2 \floor{U/2}$}
		\STATE{5: \hspace{4 mm} $\mathbf{S}^{-1}_{(i,i-1)} = - \mathbf{G}_{(i,i-1)}\mathbf{S}^{-1}_{i,i}\mathbf{S}^{-1}_{(i-1,i-1)}$}
		\STATE{6: \hspace{4 mm} $\mathbf{S}^{-1}_{(i,i+1)} = - \mathbf{G}_{(i,i+1)}\mathbf{S}^{-1}_{i,i}\mathbf{S}^{-1}_{(i+1,i+1)} $}
		\STATE{7: $\textbf{end}$}
	\end{algorithmic}
\end{algorithm}
An iterative detection method based on stair matrix is proposed in~ \cite{stair_matrix1}. In their proposed method, the initial solution is based on inverse of stair matrix and matched filter. \begin{equation}
\hat{\mathbf{x}}_{(0)} =\mathbf{S}^{-1}\hat{\mathbf{x}}_{\text{MF}}.
\label{Eq. 5}
\end{equation}
This solution is updated with each iteration in the following way:
\begin{equation}
\hat{\mathbf{x}}_{t} = \mathbf{S}^{-1}\left ( (\mathbf{S} - \mathbf{G})\hat{\mathbf{x}}_{t-1} + \hat{\mathbf{x}}_{\text{MF}}  \right )
\label{Eq. 22}
\end{equation}
It is evident that the solution only requires an inversion of the stair matrix, which is already presented in Algorithm \ref{algo1}. The rest of the algorithm requires a matrix inversion and addition with matched filter. The stair matrix based detection method requires a total of $U$ number of divisions and $t(4U^2-2U)$ number of real multiplications.
In Table \ref{tab_complexity}, the complexity of the stair matrix based detection algorithm is compared to three most popular approximate inversion based detection methods. The comparison is based on the number of real multiplication required for each detection mechanism. It is evident from the table that stair matrix based detection method is less complex than most approximate inversion based detectors.

\begin{table}[t]
	\centering
	\caption{Complexity comparison}
	\renewcommand\arraystretch{1.6}
	\label{tab_complexity}
	\begin{tabular}{|l|l|}
		\hline
		Algorithm & Computational complexity \\
		\hline
		CG &
		$(K+1)(4U^2+20U$)  \\
		\hline
		NSA &
		$(K-1)(2U^3+2U^2-2U)$ \\
		\hline
		GS &
		$6KU^2$   \\
		\hline
		Stair Matrix &
		$K(4U^2-2U)$ \\
		
\hline	
\end{tabular}
\end{table}

\section{Simulation Results}\label{simulation}

\begin{figure}[h]
\centering
\includegraphics[keepaspectratio,width=1\columnwidth]{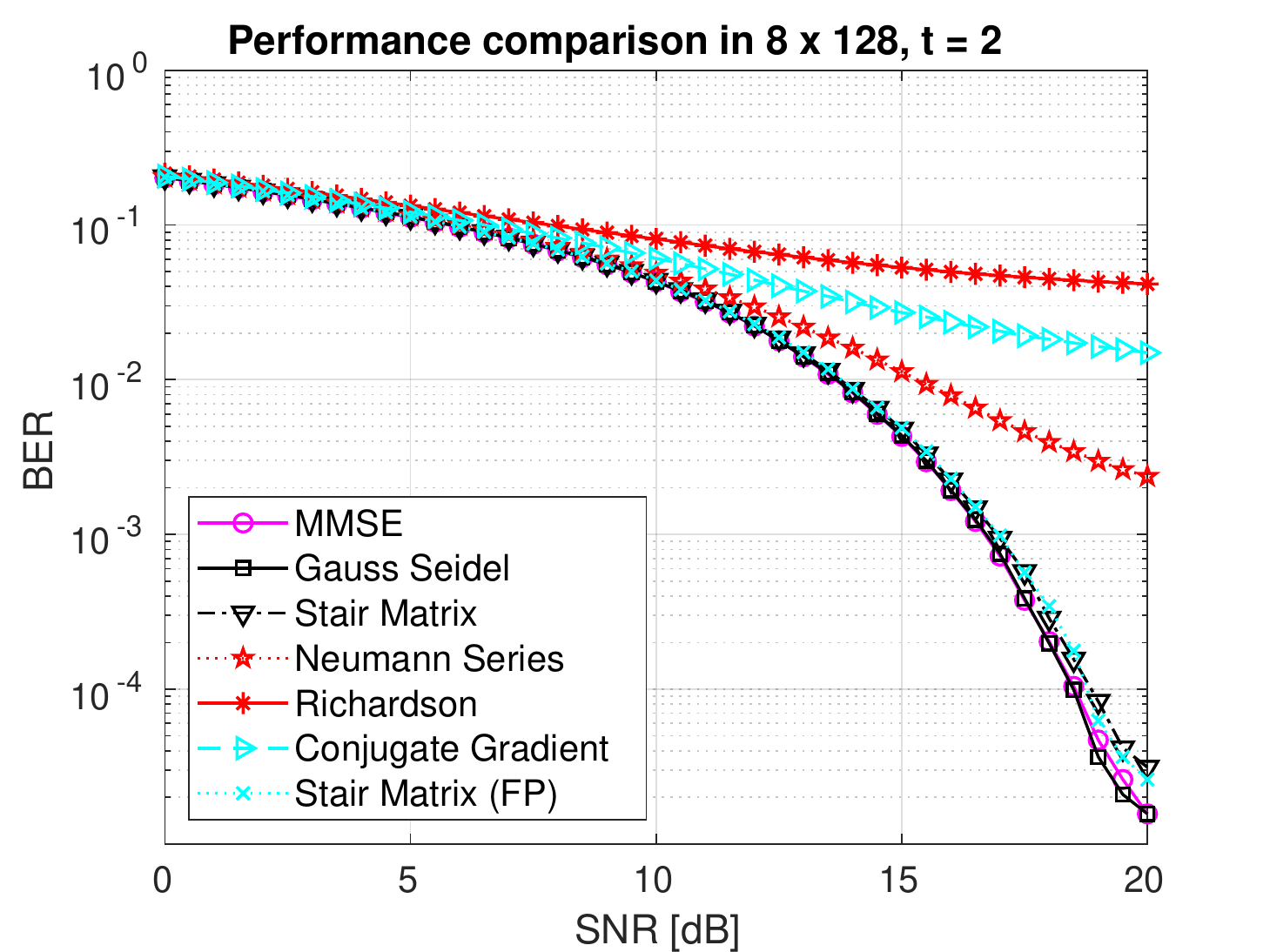}
\caption{Detector performance for 128 BS antennas and 8 users with 256-QAM.}
\label{fig128}
\end{figure}

We present error-rate performance of different detection mechanism in Fig.~\ref{fig128}. We simulate NSA, GS, MMSE, CG and Richardson methods to compare with the stair matrix based detection. Here, we used 10,000 Monte-Carlo trials for all simulations. The modulation scheme for these simulations is 256-QAM. We consider an i.i.d. Rayleigh fading channel between the BS and users.
In Fig.~\ref{fig128}, we simulate the detectors for 8 users transmitting to a BS equipped with 128 antennas. The number of iterations used for all the algorithms is $t=2$. GS and stair matrix based detectors reach near MMSE performance for this scenario. The stair matrix based detector outperforms NSA, Richardson and CG methods by a significant margin.  

We also present the fixed-point simulation performance of stair matrix in Fig.~\ref{fig128}. We curve of the stair matrix based detector with fixed word lengths is denoted by FP. The Gramian matrix is set to a total of 12 bits with 8 bits for fraction, while the matched filter inputs are set to a total of 15 bits with 10 bits of fraction. The output of the inverse of $\mathbf{S}$ is set to 17 bits. The outputs of multiplication between $(\mathbf{S}-\mathbf{G})$ and $\hat{\mathbf{x}}$ are set to 20 bits with 16 bits of fraction. The values of $\hat{\mathbf{x}}$ are set to 12 bits with 8 bits of fraction in the entire simulation. These optimal word lengths are found after numerous simulations in the Matlab environment. Wrapping and rounding mechanisms are used in the simulations for quantizing the integer and fractional parts, respectively. Fig.~\ref{fig128} shows that the fixed-point version of a stair matrix based detector coincides with its floating-point counterpart.    

\section{VLSI Architecture}\label{vlsi}
We present an iterative and time-shared VLSI architecture for the stair matrix based massive MIMO detection. The VLSI architecture is designed for maximum utilization of a complex multiplier array, which is the crux of the detector operation. A VLSI architecture based on a systolic array with numerous multiply-and-accumulate (MAC) processing elements (PE) is another candidate for such a detector. However, we prefer the proposed time-shared architecture due to the iterative nature of our algorithm. 

A high level block diagram of the proposed VLSI architecture is presented in Fig.~\ref{vlsi}. We assume the \textit{Gramian} and matched filters are computed during pre-processing stage and we focus only on the detection part in this paper. The inputs of the architecture comes from subtraction of the stair matrix from the \textit{Gramian} matrix ($\mathbf{S}-\mathbf{G}$), diagonal elements of ($\mathbf{G}$), non-diagonal elements ($\mathbf{G}$), and matched filter. A 13 bits input where 9 bits are fraction is used for $\mathbf{G}$ matrix. Even though 12 bits are sufficient for $\mathbf{G}$ according to Fig.~\ref{fig128}, we relax the word length by one bit to avoid any unexpected quantization errors.
The memory part is depicted in green, while the logic parts in blue in Fig.~\ref{vlsi}. It should be noted that the size of the blocks in Fig.~\ref{vlsi} is not proportional to the actual area they occupy in the FPGA die.

\begin{figure}[h]
\centering
\includegraphics[keepaspectratio,width=1\columnwidth]{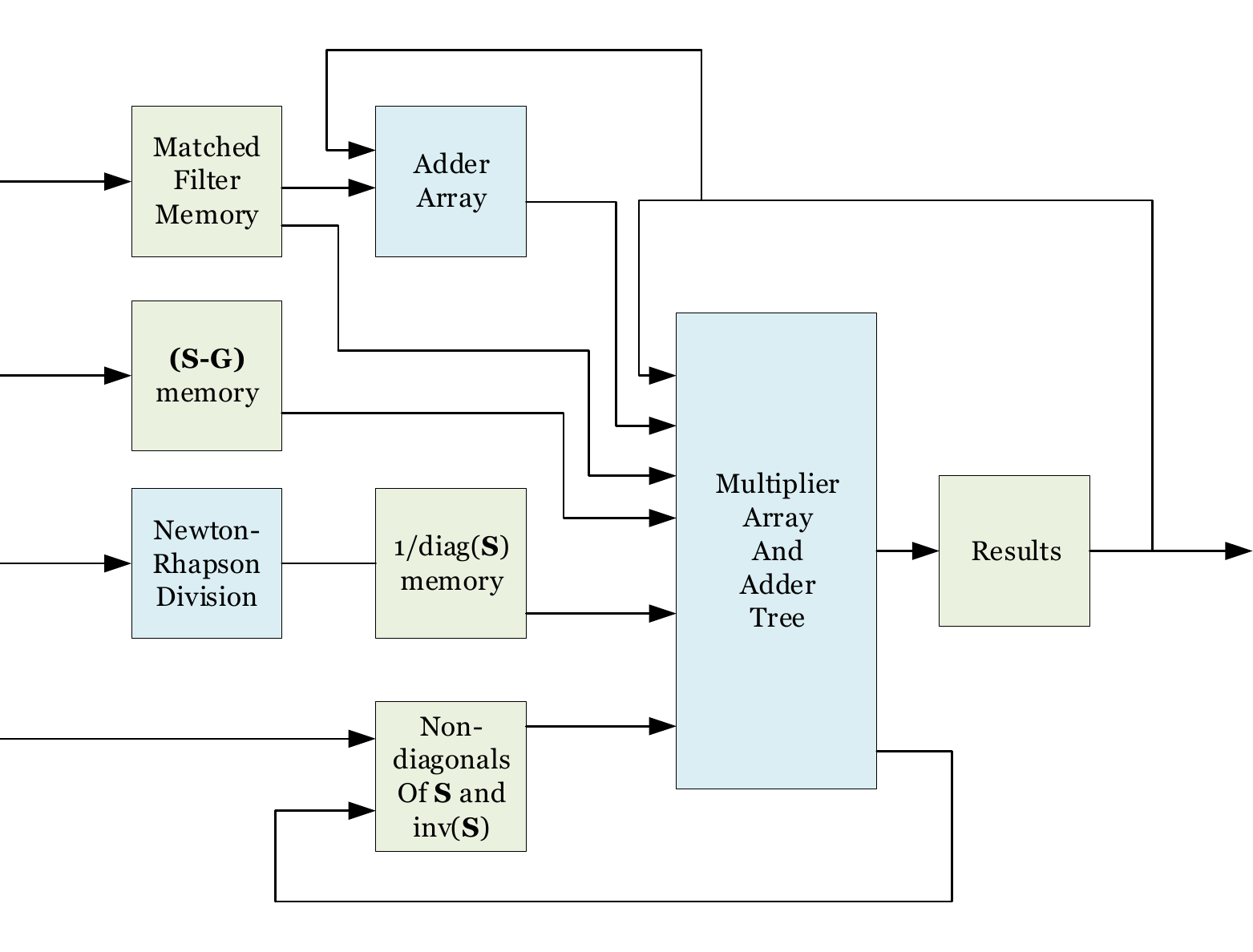}
\caption{High level architecture of the VLSI architecture for stair matrix based detection.}
\label{vlsi}
\end{figure}

According to Fig.~\ref{fig128}, the entire architecture can be divided in four major parts: (1) a Newton-Raphson divider, (2) a multiplier array, (3) an adder tree and an adder array, and (4) control logic. 
The main memories involved with the architecture is related to $\mathbf{S}-\mathbf{G}$, and register arrays to store incoming matched filter, diagonal and non-diagonals of $\mathbf{S}^{-1}$, and an array at the output. 
We require to store 64 elements of $\mathbf{S}-\mathbf{G}$ because the Gramian becomes $8\times 8$ for an 8 users massive MIMO system. We use a memory which stores 8 elements of $\mathbf{S}-\mathbf{G}$ in each address to utilize the multiplier array properly. Thus, each word in the memory is $26\times 8$ bits wide.
However, the values of $\mathbf{S}-\mathbf{G}$ is assumed to be written one at a time, and therefore, we require a separate register array and associated logic to concatenate the 8 elements into a longer word of $26\times 8$ bits and write into the $\mathbf{S}-\mathbf{G}$ memory. A few points to be noted here: (1) $\mathbf{S}-\mathbf{G}$ do not require any subtraction, because the non-zero elements of $\mathbf{S}$ comes straight away from $\mathbf{G}$, and the subtraction output will be zero for those elements. (2) Due to the symmetric nature, it is possible to store only the upper or lower triangular part of $\mathbf{G}$. However, writing and reading from a triangular memory will require more complex logic. (3) We assume the chip select, write enable, write address and write data comes as an input to the architecture.

During first 64 cycles required to load the $\mathbf{S}-\mathbf{G}$ memory, we compute the inverse of the diagonal elements with Newton-Raphson method~\cite{LR_ISCAS}. The inverse $1/x$ can be computed in an iterative manner with Newton-Raphson method as
\begin{equation}
x_{k+1} = 2x_k - xx_k^2,
\label{Eq. 22}
\end{equation}
where $k$ is the number o iterations. The initial value $x_0$ follows the convergence criterion $0 < x_0 < 2/x$. The initial values are typically stored in a look-up table (LUT). To control the dynamic range, a shift operation is applied on $x$. This shift operation ensures that the result lies in the range of $[1/2,1]$. We use a 18 bit divider and store $\text{diag}(\mathbf{G})$ in a register array.

The non-diagonal elements of $\mathbf{S}$ are stored in a separate register array. To our convenience, we only have a total of seven non-zero non-diagonal values of in a $8\times 8$ stair matrix. The corresponding non-zero elements of $1/\mathbf{S}$ can be computed with only two multiplications according to Algorithm \ref{algo1}.
We store the non-diagonal elements of $1/\mathbf{S}$ in a separate register array. 
The main computation unit of the architecture is built upon an array of eight complex multipliers and an adder tree. As complex multipliers are costly, we re-use them in a time shared manner to compute (1) initial value of $\hat{\mathbf{x}}$, (2) matrix $(\mathbf{S}-\mathbf{G})$ and vector $\hat{\mathbf{x}}$ multiplication, (3) to compute non-diagonal elements of $1/\mathbf{S}$ and (4) the final multiplication required between $\mathbf{S}^{-1}$ and $\left ( (\mathbf{S} - \mathbf{G})\hat{\mathbf{x}}_{t-1} + \hat{\mathbf{x}}_{\text{MF}}  \right )$.

\section{FPGA Implementation and Comparison}\label{fpga}
The VLSI architecture is developed with VHDL on register-transfer level (RTL) and mapped on a Xilinx Virtex-7 XC7VX690T FPGA. For synthesis and implementation strategy, Vivado default settings is used. The default mode is selected for the \texttt{-flatten\_hierarchy} option in Vivado design tool to keep the same top level hierarchy after synthesis. A total of 116 cycles are required to complete the two iterations of the stair matrix based detector, where 25 cycles are required for each iteration of the algorithm. The architecture can reach 258~MHz of maximum clock frequency. Therefore, the 8 users system can reach upto 142.34~Mbps throughput. Several such VLSI circuits working in parallel on different orthogonal frequency division multiplexing (OFDM) tone can increase the throughput to Gbps, which is required for 5G systems. 

In Table \ref{tab2}, we compare our FPGA implementation with NSA implementation of~\cite{NSA_Chris_Rice}~and GS implementations of~\cite{chuan_zhang_GS} and \cite{improved_GS}. The NSA detector takes a significant amount of LUT and FF slices compared to other implementations and able to reach a high throughput value even after $t=3$ iterations. As the implementation also takes a significant area, the NSA implementations provide lower scaled throughput compared to our implementation. The GS implementations are both provided for $t=1$ iterations. The GS implementation of~\cite{chuan_zhang_GS} provides lower throughput as well as lower scaled throughput compared to our implementation. The throughput will be further reduced for $t=2$ iterations which we have used in our Matlab simulations of Fig.~\ref{fig128}. The improved GS architecture proposed in~\cite{improved_GS} provides higher scaled throughput than that of~\cite{chuan_zhang_GS}. However, the scaled throughput is still lower than our implementation and~\cite{improved_GS} also used only $t=1$ iterations for their result. 

\section{Conclusion}\label{Conclusion}

In this paper, we proposed a VLSI architecture and FPGA implementation of a stair matrix based massive MIMO detection algorithm. The algorithm provides satisfactory performance when the ratio between the numbers of BS antennas and users is high. We presented the fixed-point simulations to find the optimal word lengths for our architecture. The implementation reaches a maximum clock frequency of 258~MHz and provides 142.34 Mbps. The architecture provides reasonably higher throughput and better error-rate performance than most other massive MIMO detectors. Thus, the proposed detector is an attractive solution for 5G BS receiver implementation..

\begin{table}[t]
\begin{threeparttable}

\caption{Comparison of FPGA implementations for 8 users massive MIMO detectors}
\renewcommand\arraystretch{1.1}
\label{tab2}
\centering
\begin{tabular}{lllll}
\toprule
Detection algorithm                 & Stair Matrix             & NSA~\cite{NSA_Chris_Rice}    & GS~\cite{chuan_zhang_GS} & GS~\cite{improved_GS}\\
Modulation Scheme                   & 256-QAM            & 64-QAM                        & 64-QAM & 64-QAM \\
Iteration                           & 2                 & 3                        & 1 & 1\\
\midrule
LUT slices          & 16211             & 148797            & 18976   & 17944  \\
FF slices            & 15793            & 161934            & 15864   & 19750\\
DSP                  & 112              & 1016              & 232    & 274\\
\midrule
Clock freq. [MHz]    &  258              & 317               & 309   &  352\\
Throughput [Mbps] &  142.34           & 621            & 48   & 127\\
\midrule
\begin{tabular}{@{}l@{}}Throughput/slices$^{a}$ \\ 
			(Mbps/K slices) \end{tabular}& 4.4481  & 2.0032 & 1.4118 & 3.4324\\
\bottomrule
\end{tabular}
\begin{tablenotes}[l]
\item[$a$]Summation of LUT and FF slices.
\end{tablenotes}

\end{threeparttable}
\end{table}

\ifCLASSOPTIONcaptionsoff
  \newpage
\fi

\bibliographystyle{IEEEtran}
\bibliography{references}

\begin{thebibliography}{1}
\providecommand{\url}[1]{#1}
\csname url@samestyle\endcsname
\providecommand{\newblock}{\relax}
\providecommand{\bibinfo}[2]{#2}
\providecommand{\BIBentrySTDinterwordspacing}{\spaceskip=0pt\relax}
\providecommand{\BIBentryALTinterwordstretchfactor}{4}
\providecommand{\BIBentryALTinterwordspacing}{\spaceskip=\fontdimen2\font plus
\BIBentryALTinterwordstretchfactor\fontdimen3\font minus
  \fontdimen4\font\relax}
\providecommand{\BIBforeignlanguage}[2]{{%
\expandafter\ifx\csname l@#1\endcsname\relax
\typeout{** WARNING: IEEEtran.bst: No hyphenation pattern has been}%
\typeout{** loaded for the language `#1'. Using the pattern for}%
\typeout{** the default language instead.}%
\else
\language=\csname l@#1\endcsname
\fi
#2}}
\providecommand{\BIBdecl}{\relax}
\BIBdecl

\bibitem{MIMO_survey}
M.~A. {Albreem}, M.~{Juntti}, and S.~{Shahabuddin}, ``Massive {MIMO}
  {D}etection {T}echniques: A {S}urvey,'' \emph{IEEE Communications Surveys
  Tutorials}, vol.~21, no.~4, pp. 3109--3132, 2019.

\bibitem{NSA_Chris_Rice}
M.~Wu, B.~Yin, G.~Wang, C.~Dick, J.~Cavallaro, and C.~Studer, ``Large-scale
  {MIMO} detection for {3GPP LTE}: Algorithm and {FPGA} implementation,''
  \emph{IEEE Journal of Selected Topics in Signal Processing}, vol.~8, no.~5,
  pp. 916--929, Oct. 2014.

\bibitem{chuan_zhang_GS}
Z.~Wu, C.~Zhang, Y.~Xue, S.~Xu, and X.~You, ``{Efficient architecture for
  soft-output massive {MIMO} detection with {Gauss-Seidel} method},'' in
  \emph{IEEE International Symposium on Circuits and Systems}, May 2016, pp.
  1886--1889.

\bibitem{MatDecomposition_mMIMO}
S.~Shahabuddin, M.~S. Islam, M.~S. Shahabuddin, M.~A. Albreem, and M.~Juntti,
  ``Matrix {D}ecomposition for {M}assive {MIMO} {D}etection,'' in \emph{IEEE
  Nordic Circuits and Systems Conference}, Oct. 2020.

\bibitem{Richardson}
X.~Gao, L.~Dai, C.~Yuen, and Y.~Zhang, ``Low-complexity {MMSE} signal detection
  based on {R}ichardson method for large-scale {MIMO} systems,'' in \emph{IEEE
  Vehicular Technology Conference}, 2014, pp. 1--5.

\bibitem{stair_matrix1}
F.~{Jiang}, C.~{Li}, Z.~{Gong}, and R.~{Su}, ``Stair matrix and its
  applications to {M}assive {MIMO} uplink data detection,'' \emph{IEEE
  Transactions on Communications}, vol.~66, no.~6, pp. 2437--2455, 2018.

\bibitem{LR_ISCAS}
S.~{Shahabuddin}, J.~{Janhunen}, Z.~{Khan}, M.~{Juntti}, and A.~{Ghazi}, ``A
  customized lattice reduction multiprocessor for {MIMO} detection,'' in
  \emph{IEEE International Symposium on Circuits and Systems (ISCAS)}, 2015,
  pp. 2976--2979.

\bibitem{improved_GS}
J.~{Zeng}, J.~{Lin}, and Z.~{Wang}, ``An improved {G}auss-{S}eidel algorithm
  and its efficient architecture for massive {MIMO} systems,'' \emph{IEEE
  Transactions on Circuits and Systems II: Express Briefs}, vol.~65, no.~9, pp.
  1194--1198, 2018.

\end{thebibliography}

\end{document}